\documentclass[aps,prl,reprint,groupedaddress,longbibliography,nofootinbib]{revtex4-1}
\usepackage{graphicx,amsmath,amssymb,braket,siunitx}
\usepackage{dcolumn}
\usepackage{bm}
\usepackage{hyperref}
\usepackage[version=4]{mhchem}
\DeclareSIUnit\gauss{G}
\DeclareSIUnit\sig{\mbox{$\sigma$}}
\usepackage{placeins}
\usepackage[colorinlistoftodos]{todonotes}

\begin{document}
\title{Microwave trap for atoms and molecules}
\author{S. C. Wright}
\author{T. E. Wall}
\author{M. R. Tarbutt}
\affiliation{Centre for Cold Matter, Blackett Laboratory, Imperial College London, Prince Consort Road, London SW7 2AZ UK
}

\begin{abstract}
We demonstrate a trap that confines polarizable particles around the antinode of a standing-wave microwave field. The trap relies only on the polarizability of the particles far from any resonances, so can trap a wide variety of atoms and molecules in a wide range of internal states, including the ground state. The trap has a volume of about \SI{10}{\centi\metre\cubed}, and a depth approaching \SI{1}{\kelvin} for many polar molecules. We measure the trap properties using $^{7}$Li atoms, showing that when the input microwave power is \SI{610}{\watt}, the atoms remain trapped with a $1/e$ lifetime of \SI{1.76 (12)}{s}, oscillating with an axial frequency of \SI{28.55 (5)}{\hertz} and a radial frequency of \SI{8.81 (8)}{\hertz}. The trap could be loaded with slow molecules from a range of available sources, and is particularly well suited to sympathetic cooling and evaporative cooling of molecules.
\end{abstract}


\maketitle

Almost all research using cold atoms, molecules and ions relies on trapping. The trap confines the particles to a small volume so that they can be cooled to low temperature, collide with one another, and be studied and controlled with high precision. Thus, the trap is a key tool in frequency metrology, quantum information processing, quantum simulation, field sensing, cavity quantum electrodynamics, studies of quantum degenerate gases, tests of fundamental physics, and many other topics. New traps often stimulate new applications, and new research areas often call for new traps. Here, we demonstrate a trap that confines particles around the electric field antinode of a standing-wave microwave field formed inside an open resonator, realizing the proposal of DeMille et al.~\cite{DeMille2004}. The trap relies only on the polarizability of the particles at microwave frequencies, far from any resonances, so is suitable for trapping a wide variety of atoms and molecules.

For atoms, the microwave trap has a depth similar to an optical dipole trap, but its volume is a million times greater so it can trap samples with a much lower phase-space density. Importantly, heating due to spontaneous emission, which often limits the lifetime of an optical trap, is eliminated in the microwave trap. The usefulness of a microwave trap for atoms was recognized long ago~\cite{Agosta1989} in the context of evaporative cooling to quantum degeneracy. Such a trap was developed specifically for ultracold Cs~\cite{Spreeuw1994}, but it used the magnetic dipole interaction at a frequency almost resonant with the ground-state hyperfine transition, so was specific to that particular atom. More recently, a similar species-specific microwave-induced force has been used to generate spin-dependent potentials on atom chips, where strong gradients can be produced in the near-field of coplanar waveguides and resonators~\cite{Treutlein2006, Bohi2009, Fancher2018}. By contrast, ours is a very general trap that uses the electric dipole interaction far from any resonance.

Especially important at present is the development of new traps for cold, polar molecules, which can be used to test fundamental physics~\cite{Hudson2011, Baron2014, Cairncross2017, Andreev2018, Hudson2006, Truppe2013, Altuntas2018, Hunter2012}, study cold chemistry~\cite{Krems2008, Richter2015, Liu2018}, process quantum information~\cite{DeMille2002, Yelin2006, Andre2006}, and explore interacting many-body quantum systems~\cite{Micheli2006, Barnett2006, Gorshkov2011, Yan2013}. Some molecular species can now be formed at sub-millikelvin temperatures by direct laser cooling~\cite{McCarron2018b, Tarbutt2019}, optoelectrical cooling~\cite{Prehn2016}, or by association of ultracold atoms~\cite{Moses2017, DeMarco2018}, and they have been confined in magnetic traps~\cite{Williams2018, McCarron2018}, electric traps~\cite{Prehn2016}, and optical traps~\cite{Danzl2010, Chotia2012, Anderegg2018}. A wider variety of molecules can be produced in the 10-100~mK range using a set of techniques that includes buffer-gas cooling, Stark, Zeeman and centrifuge deceleration~\cite{Weinstein1998, Bethlem1999, Hogan2008, Chervenkov2014}. These warmer molecules could be sympathetically cooled to much lower temperatures through collisions with co-trapped ultracold atoms~\cite{Tokunaga2011,Lim2015}. This requires trapping of ground-state molecules so that inelastic collisions that inhibit sympathetic cooling are energetically forbidden. Unfortunately, ground-state particles are always strong-field-seeking, so cannot be confined in static electric and magnetic traps~\cite{Earnshaw1842}. One possible solution is the ac electric trap~\cite{vanVeldhoven2005}, whose operating principle is similar to that of a Paul trap for charged particles. However, this method suffers from a small trap depth, typically below \SI{10}{\milli\kelvin}, and a small volume of around \SI{e-2}{\centi\metre\cubed}, and is not compatible with sympathetic cooling~\cite{Tokunaga2011}. Optical dipole traps can also trap ground-state molecules, but usually have depths below \SI{1}{\milli\kelvin} and volumes of about \SI{e-5}{\centi\metre\cubed}. By contrast, the microwave trap has a volume of about \SI{10}{\centi\metre\cubed} and a depth in the range 0.1--1~K for many polar molecules. It has previously been shown that microwave fields in high quality-factor resonators can be used to deflect or focus beams of NH$_3$~\cite{Odashima2010}, CH$_3$CN~\cite{Spieler2013} and PbO~\cite{Enomoto2019}, and to decelerate a beam of NH$_3$ by a few m/s~\cite{Merz2012}. However, atoms and molecules have never previously been trapped this way. Using ultracold $^7$Li, we show that the trap works and we measure its properties.

\begin{figure}[t]
    \centering
    \includegraphics[width = 0.5\textwidth]{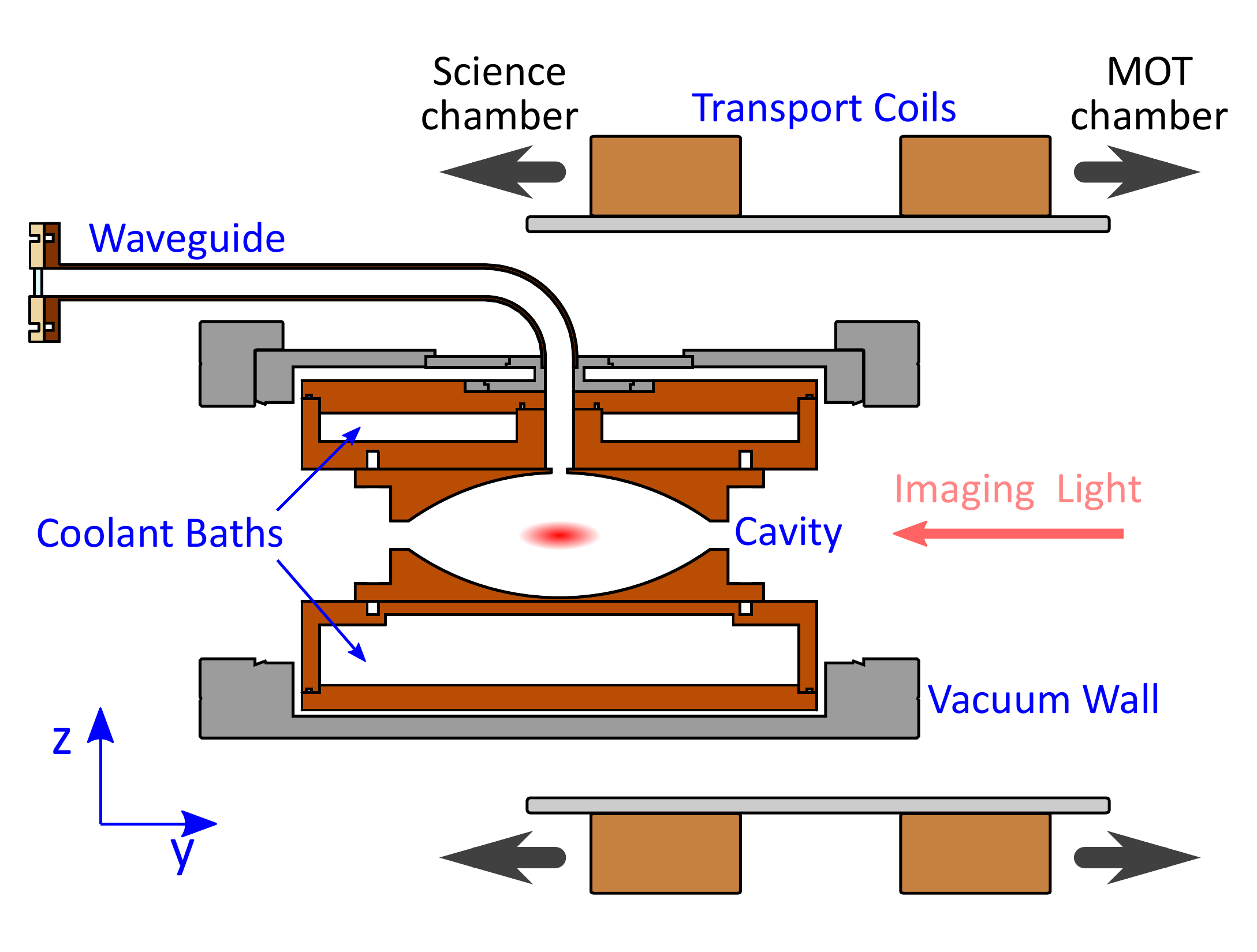}
    \caption{Illustration of the experiment, showing the microwave trap and the transport coils.}
    \label{fig:ExptSetup}
\end{figure}

Figure \ref{fig:ExptSetup} illustrates the microwave trap and the moving magnetic trap that delivers the atoms. The design of the microwave trap follows that of Ref.~\cite{Dunseith2015}. Two copper mirrors, cooled using water flowing at \SI{0.5}{\litre\per\minute}, form a Fabry-P\'erot cavity. The mirrors have diameter 90~mm, radius of curvature $R_{\rm m} = $ \SI{73}{mm}, and a center-to-center separation of $L=$ \SI{35}{mm}. We use the lowest-order Gaussian mode with longitudinal mode index $n=3$ (TEM$_{003}$), whose resonant frequency, $f_{n}$, is near \SI{14.27}{\giga\hertz}. For this mode, the unloaded quality factor at room temperature is $Q_0 \approx$ \SI{2e4}{}. Power is coupled into the cavity through a hole, of diameter $d_h=$ \SI{4.63}{mm} and thickness $t=\SI{0.7}{mm}$, at the centre of one mirror. For this choice of hole size, 95\% of the incident power is transmitted into the cavity. The perfect Gaussian mode would have a beam waist of $w_0 =\SI{14.7}{mm}$, but the coupling hole broadens the mode, and we measure $w_0 = \SI{17.25}{mm}$.  To feed the cavity, the signal from a microwave oscillator is amplified by a klystron, which provides \SI{80}{dB} gain and a maximum output power of \SI{2}{kW}. The power is delivered via a waveguide which interfaces directly with the coupling hole. A window in the waveguide flange seals the vacuum. A sinusoidal frequency modulation of amplitude \SI{40}{kHz} and frequency \SI{20}{kHz} is applied to the oscillator. Directional couplers pick off \SI{-40}{dB} of the incident and reflected powers, and the ratio of these signals is used as the input to a lock-in amplifier which locks the microwave frequency to the cavity resonance by minimizing the reflected power. We define $P$ to be the incident power transmitted into the cavity, and determine its value by measuring the power output from the klystron and accounting for the fraction absorbed by the waveguide and reflected by the cavity. The electric field amplitude at the centre of the cavity is
\begin{equation}
\mathcal{E}_0 = \left(\frac{4 P Q_0}{\pi^2 \epsilon_0 f_n w_0^2 L}\right)^{1/2}. 
\end{equation}
When $P=$ \SI{700}{W}, $\mathcal{E}_0 \approx $ \SI{20}{\kilo\volt\per\centi\metre}. For Li, whose static scalar polarizability is $\alpha_{\rm s} = \SI{2.70e-39}{\joule\metre^2\per\volt^2}$~\cite{Molof1974}, the corresponding trap depth is $U_0 = \alpha_{\rm s}  \mathcal{E}_0^2/4 \approx \SI{200}{\micro\kelvin}$. The magnetic field in the cavity couples off-resonantly to the ground state hyperfine structure, but this provides a correction of less than $1.5\%$ to the  trap potential.     

Each experiment begins by loading \SI{1e8}{} $^7$Li atoms into a magneto-optical trap (MOT) at a temperature of \SI{1.07(6)}{mK}. The atoms are cooled further, to \SI{50}{\micro\kelvin}, using Raman gray molasses on the D$_1$ line~\cite{Grier2013}, then optically pumped into the $|F=2, m_F = 2 \rangle$ state. These atoms are trapped in a magnetic quadrupole trap using coils inside the MOT vacuum chamber, which produce an axial field gradient of \SI{32}{\gauss\per\centi\metre}. Then, they are transferred adiabatically to a second quadrupole trap, formed by coils external to the vacuum chamber and mounted on a motorized translation stage. The axial field gradient is ramped up to \SI{50}{\gauss\per\centi\metre} and then the trap is translated horizontally by \SI{600}{mm}, bringing the atoms to the center of the microwave trap, which is housed in a separate vacuum chamber from the MOT. At this point the magnetic trap contains about $2 \times 10^7$ atoms, at a phase-space density of $3.6(2) \times 10^{-7}$. Next, the microwave power is ramped linearly in \SI{200}{ms} from an initial value\footnote{This small initial power is needed to maintain the frequency lock. It is insufficient to form an axial trap when the gravitational potential is included.} of \SI{10}{W} at $t=-\tau_{\rm ramp}$, to the final trapping power, $P$, at $t=0$. The magnetic trap currents are ramped down over the same period, reaching zero at $t=0$, which defines the start of the microwave trapping period. Unless stated otherwise, we use $\tau_{\rm ramp} = \SI{200}{ms}$. After a variable hold time in the microwave trap, we return the atoms to the magnetic trap and turn off the microwave trap. The density distribution of the atoms, in either the microwave trap or magnetic trap, is measured by absorption imaging using light resonant with the $F=2 \rightarrow F'= 3 $ $D_2$ transition.

\begin{figure}[t]
    \centering
    \includegraphics[width = 0.45\textwidth]{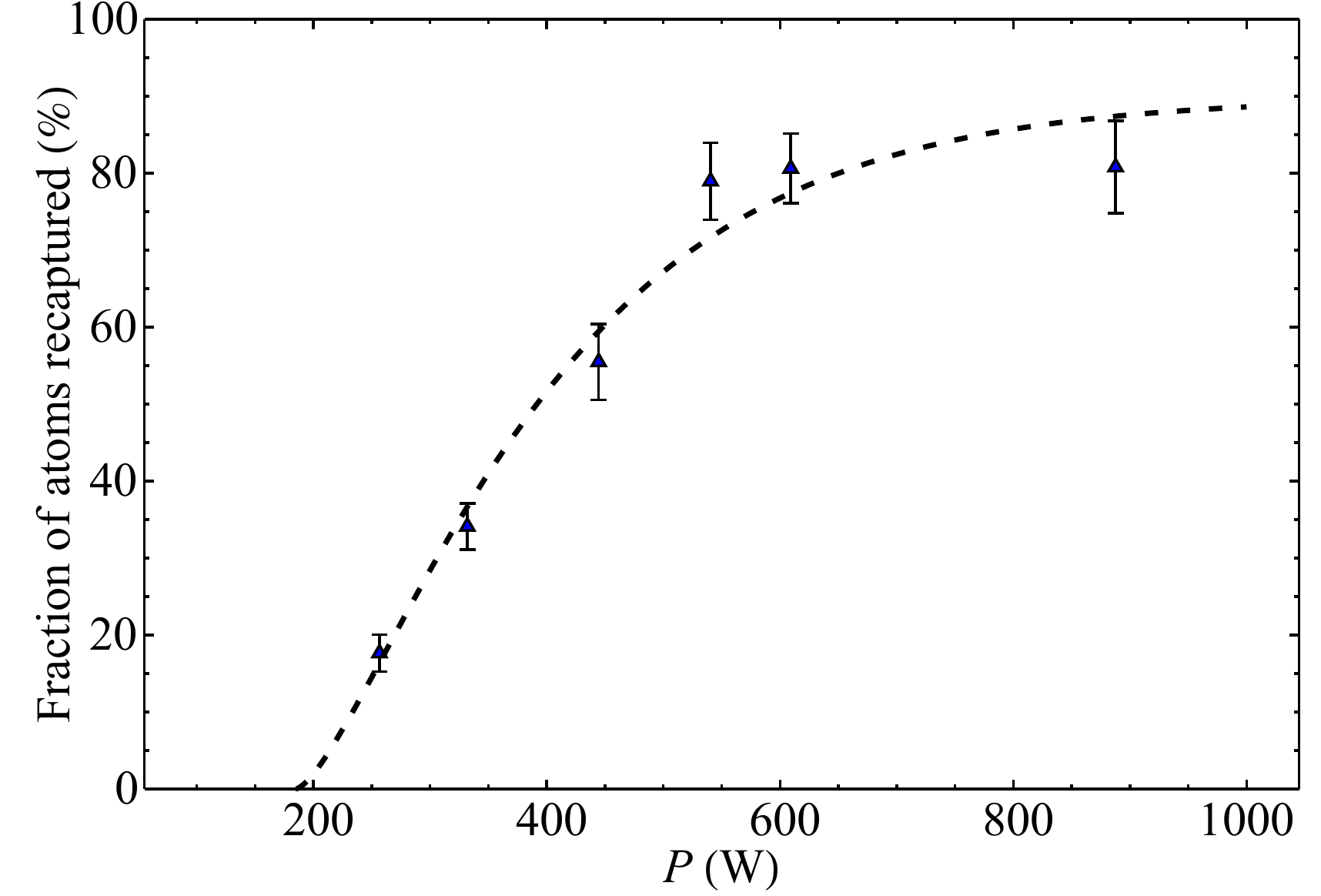}
    \caption{Fraction of atoms recaptured into the magnetic trap as a function of input microwave power, $P$. Dashed line: fit to Eq.~(\ref{eqn:RecapturePrediction}), fixing $\beta =\SI{0.28}{\micro\kelvin\per\watt}$.}
    \label{fig:RecaptureVsPower}
\end{figure}

Figure \ref{fig:RecaptureVsPower} shows the fraction of atoms recaptured into the magnetic trap at $t=\SI{200}{ms}$, as a function of $P$. At low power, we do not recapture any atoms because the sum of the microwave and gravitational potentials does not form a trap until $P$ exceeds a threshold value, $P_0$. The fraction recaptured, $\eta$, then increases with $P$, and begins to saturate at $P \approx \SI{600}{W}$. Only two of the five $F=2$ states are magnetically trappable, so no more than $40\%$ would be recaptured from a randomized spin ensemble. Since we recapture a greater fraction than this, we conclude that the spin polarization is fairly well preserved in the microwave trap due to a background magnetic field of around 1~G. This is no surprise, since the spin precession frequency is faster than the oscillation frequency in the trap, provided the background field is greater than \SI{0.02}{mG}. We fit the data in Fig.~\ref{fig:RecaptureVsPower} with the simple model
\begin{equation}
    \eta =  \eta_{\rm max} \int_{0}^{\beta(P-P_0)} \frac{2\sqrt{E}}{\sqrt{\pi}(k_B T)^{\frac{3}{2}}} \exp\left(-\frac{E}{k_B T}\right) dE.  
    \label{eqn:RecapturePrediction}
\end{equation}
Here, the integrand is the initial distribution of energies, $E$, characterized by the temperature, $T$, and we integrate this up to the trap depth, $\beta(P-P_0)$, where $\beta = \SI{0.28}{\micro\kelvin\per\watt}$ is the calculated gradient of the trap depth versus power. The best fit parameters are $T = \SI{44(12)}{\micro\kelvin}$, $\eta_{\rm  max} = 90(9)\%$, and $P_0 = \SI{180(20)}{W}$. For atoms loaded exactly at the antinode of the microwave field, the calculated threshold power is \SI{150}{W}. The fitted value is consistent with loading the trap about \SI{1}{mm} too high.

\begin{figure}[tb]
    \centering
    \includegraphics[width=0.5\textwidth]{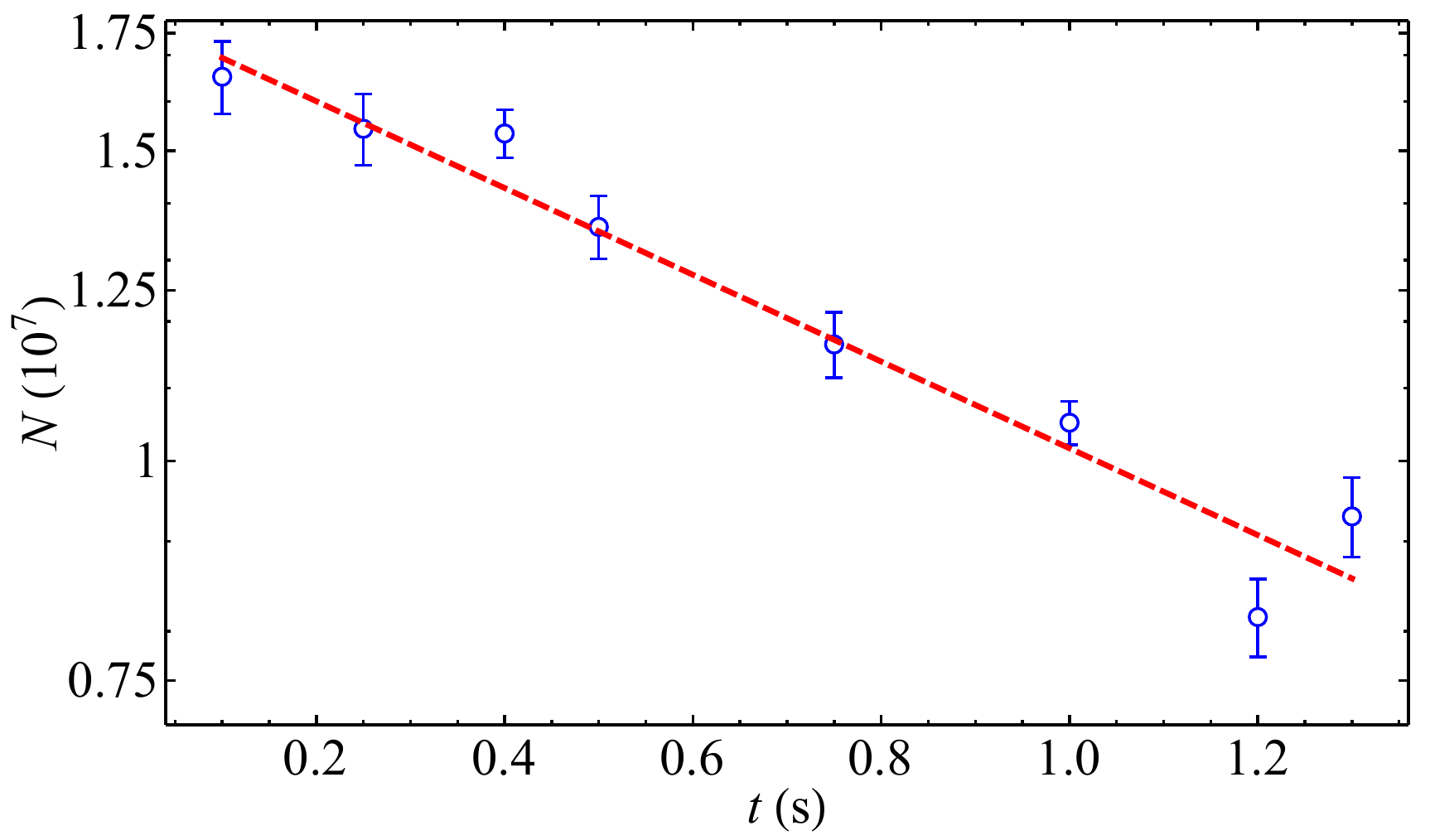}
    \caption{Number of atoms recaptured into the magnetic trap (shown on a logarithmic scale) versus time in the microwave trap. Dashed, red line: fit to $N(t)=N_0 e^{-t/\tau}$, giving $\tau = $ \SI{1.76(12)}{s}.}
    \label{fig:MWlifetimeFigure}
\end{figure}

Figure \ref{fig:MWlifetimeFigure} shows the number of atoms recaptured into the magnetic trap as a function of hold time in the microwave trap, with $P=$ \SI{610}{W}. For $0<t<$ \SI{1.3}{s} the loss is slow, and the data fit well to an exponential decay with a $1/e$ lifetime of \SI{1.76 (12)}{s}. For these data, the pressure in the microwave trap chamber was \SI{2e-9}{mbar}. The measured lifetime is consistent with that of the magnetic trap at the same pressure, suggesting that there are no significant loss mechanisms from the microwave trap other than collisions with background gas. Beyond \SI{1.5}{s}, we see a sudden increase in the pressure, typically by a factor of 10, accompanied by a corresponding increase in the loss rate. We attribute this to outgassing by microwave-absorbing materials in the chamber, and are currently investigating this.

\begin{figure}[tb]
    \centering
    \includegraphics[width = 0.5\textwidth]{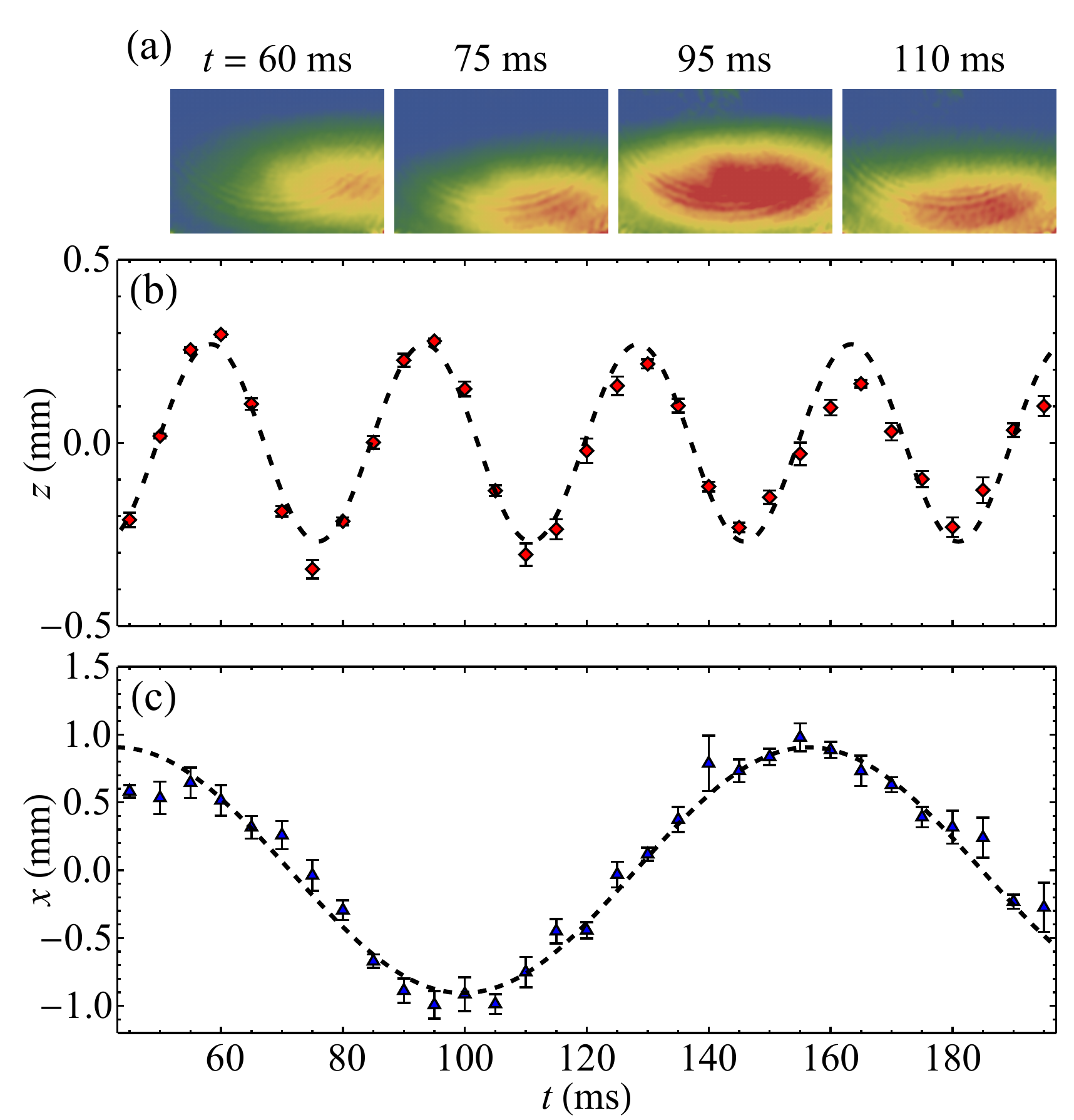}
    \caption{Oscillations following an off-centre release into the microwave trap with $P=\SI{610}{W}$. (a) Absorption images at selected times. (Horizontal and vertical directions are radial ($x$) and axial ($z$), and the image size is \SI{5.5}{mm} $\times$ \SI{3.7}{mm}). (b) Centre-of-mass motion along $z$. (c) Centre-of-mass motion along $x$. Black, dashed lines are fits to a sinusoidal model.}
    \label{fig:TrapOscillationDemo}
\end{figure}

To determine the oscillation frequencies in the trap, we release molecules more suddenly from the magnetic trap into the microwave trap by choosing $\tau_{\rm ramp} = \SI{40}{ms}$, then measure the subsequent evolution of the density distribution. Our measurements begin at $t=\SI{50}{ms}$ to allow eddy currents induced in the cavity assembly to decay. Figure \ref{fig:TrapOscillationDemo}(a) shows absorption images of the cloud in the $xz$-plane at selected times in the microwave trap, when $P = \SI{610}{W}$. Although the available field of view is limited, it is sufficient to determine both the centre and the width of the cloud by fitting to a two-dimensional Gaussian density distribution. The axial and radial rms widths, averaged over 120 images during the oscillations, are $\sigma_{z} = \SI{1.06(8)}{mm}$ and $\sigma_{x} = \SI{2.7(3)}{mm}$ respectively. Figures \ref{fig:TrapOscillationDemo}(b,c) show the axial and radial positions of the centre of the cloud as a function of time, together with fits to the model $r=r_0 + a_r \sin(\Omega_r t + \phi_r)$, with $r \in \{z,x\}$. The fits give axial and radial oscillation frequencies of $\Omega_z/(2\pi) = \SI{28.55(5)}{Hz}$ and $\Omega_x/(2\pi) = \SI{8.81(8)}{Hz}$.

\begin{figure}[tb]
    \centering
    \includegraphics[width =  0.48\textwidth]{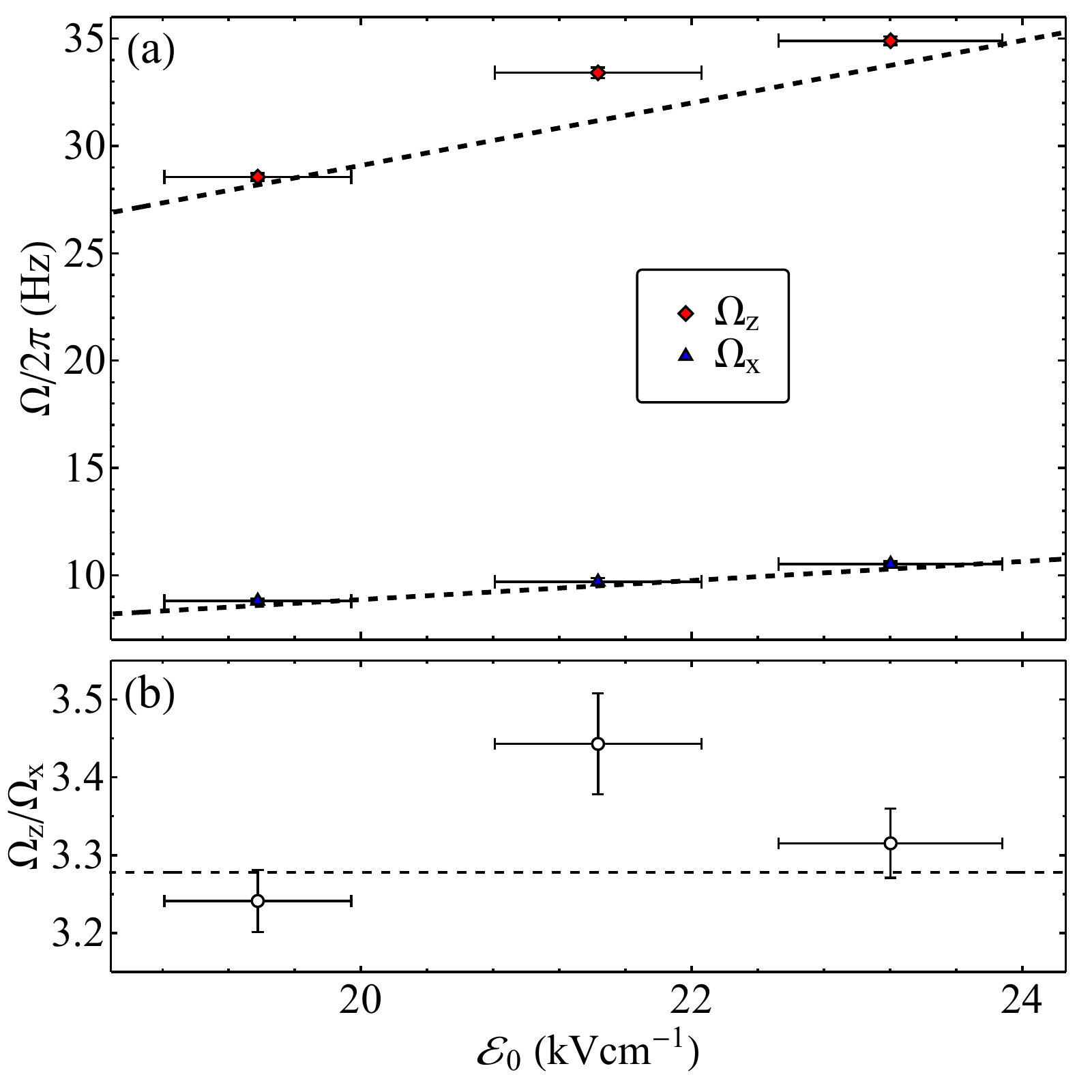}
    \caption{(a) Axial and radial oscillation frequencies as a function of the electric field at the cavity centre, $\mathcal{E}$, and (b) their ratio as a function of $\mathcal{E}_0$. Black, dashed lines: Prediction using Eq.~(\ref{eqn::TrapFrequencies}).}
    \label{fig:TrapOscillationWithTheory}
\end{figure}

By expanding the potential energy of the atoms to second order in $x$ and $z$, we find that the angular oscillation frequencies are~\cite{Wright2019}
\begin{equation}
\Omega_x  = \sqrt{\frac{\alpha_{\rm s}\mathcal{E}_{0}^2}{m w_0^2}},\hspace{0.2cm}
\Omega_z = \sqrt{\frac{\alpha_{\rm s}\mathcal{E}_{0}^2k^2(1-2\epsilon+ 2\epsilon^2)}{2 m}},
\label{eqn::TrapFrequencies}
\end{equation}
where $k$ is the wavevector, $m$ is the mass, $\epsilon = 1/(k z_0)$, and $z_0$ is the Rayleigh range. Figure \ref{fig:TrapOscillationWithTheory}(a) shows how the oscillation frequencies vary with $\mathcal{E}_0$, and compares these measurements with Eq.(\ref{eqn::TrapFrequencies}). The error bars are a measure of the systematic uncertainties in determining $\mathcal{E}_0$, which come from the uncertainties in determining $P$ and $Q_0$. Within these uncertainties, the measured frequencies are consistent with the predictions. The electric field amplitude is systematically a little higher than our estimates. Figure \ref{fig:TrapOscillationWithTheory}(b) shows the ratio $\Omega_z/\Omega_x$, which is independent of the power, as we would expect. The mean ratio is $3.33(6)$, consistent with the predicted value of $3.28$.

The atoms cool as they expand from the magnetic trap into the microwave trap. From the measured cloud sizes and trap frequencies, the relation $k_{\rm B} T = m \Omega^{2}\sigma^{2}$, and the assumption that the two radial directions are equivalent, we deduce a geometric mean temperature in the microwave trap of $T=\SI{22(3)}{\micro\kelvin}$. This is within $2\sigma$ of the temperature deduced from the fit in Fig.~\ref{fig:RecaptureVsPower}, and is a more reliable measurement. The density of atoms in the microwave trap is $\SI{1.5(3)e8}{\per\centi\metre\cubed}$, and the corresponding dimensionless phase-space density is $4(1)\times 10^{-7}$. This is consistent with the phase-space density of $3.6(2)\times 10^{-7}$ measured in the magnetic trap, implying that, within the uncertainty of 25\%, there is no loss of phase-space density in transferring atoms into the microwave trap.

\begin{figure*}[t]
    \centering
    \includegraphics[width =  \textwidth]{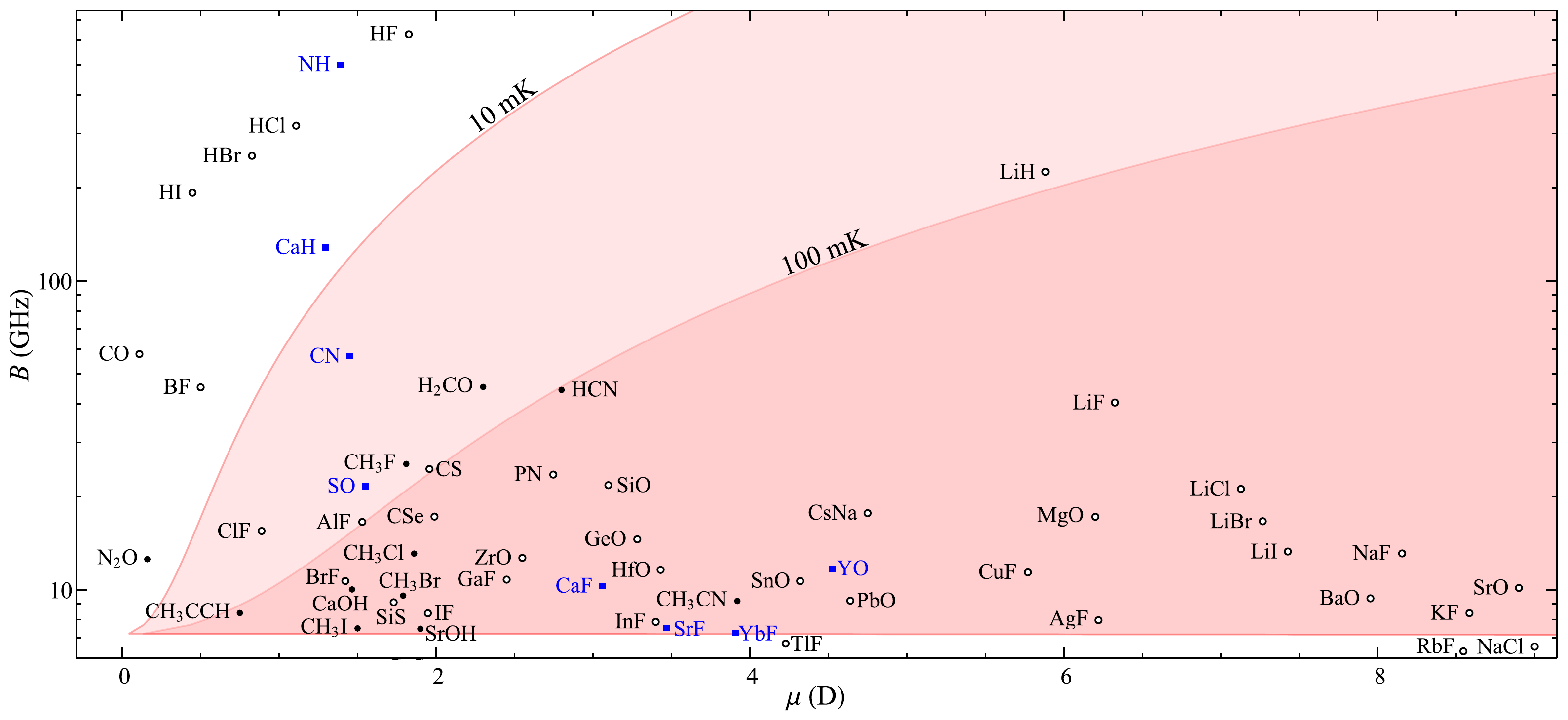}
    \caption{Rotational constants, $B$, and electric dipole moments, $\mu$, for a selection of diatomic molecules with $^{2S+1}\Sigma$ ground electronic states, and a few polyatomic molecules. Data for diatomic molecules are taken from \cite{Herzberg}; open circles are for molecules with $S=0$, and blue, solid rectangles are for those with $S > 0$. Data for polyatomic molecules (solid circles), are taken from Refs.~\cite{PolyatomicsData, CaOHandSrOHdipoles, CaOHRotConst, SrOHRotConst}. Shaded regions indicate the microwave trap depth for the parameters discussed in the text.}
    \label{fig:MolConstsPlot}
\end{figure*}

The microwave trap will work for most other laser-coolable atoms, especially the alkali and alkaline-earth atoms, whose polarizabilites are all similar to, or greater than, that of Li~\cite{Schwerdtfeger2019}. It is especially useful for laser-cooled atoms that cannot be trapped magnetically, such as Ca, Sr and Yb, and for applications where ground-state atoms are needed, where heating due to spontaneous emission must be eliminated, or where a uniform magnetic field must be applied. To overcome gravity, the heavier atoms will require a greater threshold power or the use of a magnetic field gradient to levitate the sample.

The microwave trap is very deep for a wide range of polar molecules. Figure \ref{fig:MolConstsPlot} shows a selection of molecules organised by their dipole moments, $\mu$, and rotational constants, $B$. We have selected diatomic molecules whose electronic ground states are $\Sigma$ states, along with a few experimentally relevant polyatomic molecules whose lowest rotational frequency is close to our microwave frequency. The shaded regions define the parameter space where the trap depth in our cavity would exceed \SI{10}{mK} and \SI{100}{mK}. Here, we have taken $\mathcal{E}_0 = \SI{23.2}{\kilo\volt\per\centi\metre}$, the maximum electric field strength used in our experiments, and have estimated the trap depth using perturbation theory, which gives a reasonable approximation at these depths. The figure shows 25 $^1\Sigma$ diatomic molecules (open circles) where the trap depth is greater than \SI{100}{mK}. The microwave trap is particularly useful for these species, since they cannot be trapped magnetically. A more accurate calculation of the trap depth at strong microwave fields can be obtained using the method presented in~\cite{DeMille2004}; as examples, using the electric field strength demonstrated in the present work, we estimate a trap depth of \SI{0.09}{K} for LiH, \SI{0.48}{K} for CaF, \SI{0.48}{K} for YbF, and \SI{0.65}{K} for CH$_3$CN. At higher electric fields, the trap depth may be limited by multi-photon absorption processes~\cite{DeMille2004}, but these do not become limiting until the depth is similar to the rotational constant, which is typically of order 1~K. The use of circularly-polarized microwaves avoids this problem altogether~\cite{DeMille2004}. 

Its large depth and volume make the trap suitable for capturing molecules from Stark, Zeeman or centrifuge decelerators~\cite{Bethlem1999, Osterwalder2010, Akerman2017, Wu2017}, or directly from a cryogenically-cooled buffer gas beam~\cite{Lu2014}. Because the cavity ring-down time of $\lesssim \SI{1}{\micro\second}$ permits rapid switching of the microwave field, the cavity can be used as a microwave decelerator \cite{MWStarkDecelProp, Merz2012} that brings molecules to rest at the centre of the trap. For example, CH$_3$CN molecules emerging from a bent electrostatic guide and entering our cavity along the $z$-axis with a speed of 20~m/s~\cite{Liu2010} will come to rest in two stages of microwave deceleration. Similarly, a beam of CH$_3$F molecules from a Stark or centrifuge decelerator, entering at 15~m/s~\cite{Meng2015,Chervenkov2014} will come to rest in 6 stages of microwave deceleration. 

The trap could also be used to compress samples of ultracold molecules produced by direct laser cooling, which tend to have large sizes and correspondingly low densities. For the \SI{5}{\micro\kelvin} CaF clouds recently produced~\cite{Cheuk2018, Caldwell2018}, an adiabatic compression in the microwave trap, by gradually increasing the power, would increase the density by a factor $10^{3}$. Alternatively, it could be used to implement the rapid compression method described in Ref.~\cite{Caldwell2018}, potentially increasing the density by a factor $10^{5}$. The microwave trap offers a particularly favourable environment for sympathetic cooling of molecules using ultracold atoms~\cite{Tokunaga2011, Lim2015}, or evaporative cooling of molecules, so will be an important tool for cooling a much wider range of molecules to low temperature than is currently possible.  It has been noted that the strong microwave-induced dipole-dipole interactions between molecules in the trap results in very large elastic collision cross-sections, which increase as the temperature decreases, and that this is ideal for runaway evaporative cooling of molecules~\cite{DeMille2004, Avdeenkov2012}.

Underlying data may be accessed from Zenodo\footnote{10.5281/zenodo.3237240} and
used under the Creative Commons CCZero license.
 
We thank Jon Dyne and Giovanni Marinaro for their expert technical assistance. We are grateful to Stefan Truppe, Lyra Dunseith, Ben Sauer and Ed Hinds for earlier work on the design of the experiment. S. Wright gratefully acknowledges support from the Imperial College President's PhD Scholarship scheme. This was supported by EPSRC under grants EP/I012044/1, EP/M027716/1 and EP/P01058X/1.

\bibliography{references}

\end{document}